\documentclass[aps,pre,reprint,superscriptaddress,onecolumn,longbibliography]{revtex4-2}

\pdfoutput=1
\usepackage{graphicx}
\usepackage{bm}

\usepackage{natbib}
\usepackage{hyperref}
\newcommand{\um}{$\mathrm{\mu m}$}
\newcommand{\perox}{$\mathrm{H}_2\mathrm{O}_2$}

\begin{document}


\title{Interfacial accumulation of self-propelled Janus colloids in sessile droplets}



\author{Maziyar Jalaal}
\affiliation{Van der Waals-Zeeman Institute, Institute of Physics, University of Amsterdam, Amsterdam, The Netherlands}

\author{Borge ten Hagen}
\affiliation{Physics of Fluids, University of Twente, Enschede, The Netherlands}

\author{Hai le The}
\affiliation{Physics of Fluids and BIOS Lab on a Chip, University of Twente, Enschede, The Netherlands}

\author{Christian Diddens}
\affiliation{Physics of Fluids, University of Twente, Enschede, The Netherlands}
\affiliation{Department of Physics, Eindhoven Technical University, Eindhoven, The Netherlands}

\author{Detlef Lohse}
\affiliation{Physics of Fluids, University of Twente, Enschede, The Netherlands}
\affiliation{Max Planck Center Twente for Complex Fluid Dynamics and J.M. Burgers Centre for Fluid Mechanics, University of Twente}

\author{Alvaro Marin}
\affiliation{Physics of Fluids, University of Twente, Enschede, The Netherlands}\email[]{a.marin@utwente.nl}
\homepage[]{marinlab.com}



\begin{abstract}
Living microorganisms in confined systems typically experience an affinity to populate boundaries. 
The reason for such affinity to interfaces can be a combination of their directed motion and hydrodynamic interactions at distances larger than their own size. Here we will show that self-propelled Janus particles (polystyrene particles partially coated with platinum) immersed in droplets of water and hydrogen peroxide tend to accumulate in the vicinity of the liquid/gas interface. Interestingly, the interfacial accumulation occurs despite the presence of an evaporation-driven flow caused by a solutal Marangoni flow, which typically tends to redistribute the particles within the droplet's bulk. 
By performing additional experiments with passive colloids (flow tracers) and comparing with numerical simulations for both particle active motion and the fluid flow, we disentangle the dominating mechanisms behind the observed interfacial particle accumulation. 
These results allow us to make an analogy between active Janus particles and some biological microswimmers concerning how they interact with their environment.

\end{abstract}


\maketitle


\section{Introduction} \label{Intro}

Living microorganisms are often found in confined systems, where they typically experience an affinity to populate boundaries and their vicinities. For example, the model microorganism \emph{E. coli}, a motile anaerobic bacteria, has been shown to populate the interface of small emulsion droplets. This is achieved as a result of the bacteria's directed motion, similar to a rarefied confined gas with a mean free path comparable to the confinement length scale \cite{vladescu2014ecolidroplet}. However, while this mechanism can explain the frequency of encounters with the surface, it is due to hydrodynamic interactions at distances larger than their own size \cite{frymier1995three,BergLauga2008} that \emph{E. coli} remain swimming close to such surfaces. Other swimming microorganism as \emph{B. subtilis}, a motile aerobic bacteria, tend to accumulate along the liquid/air interface of droplets due to both chemotaxis and buoyancy. At high enough concentrations, the bacterial suspension becomes hydrodynamically unstable, leading to bioconvective plume formation \cite{tuval2005bacterial}. {In a recent study, Bitterman \emph{et al.}\cite{Deblais2021algae} observed that phototactic motile algae tend to accumulate at the border of sessile droplets when illuminated under the adequate light conditions.}

\begin{figure}[b]
	\centering
	\includegraphics[width=.8\columnwidth]{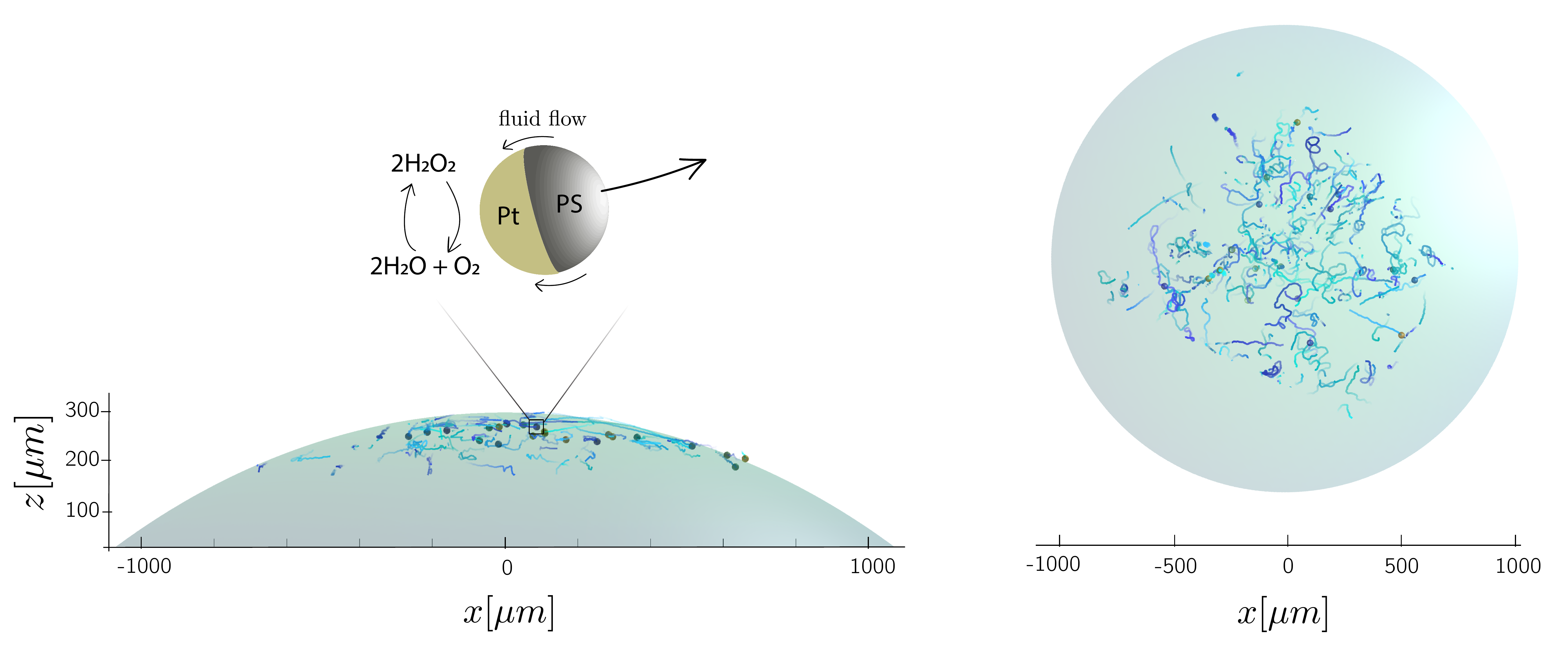}

	\caption{Active Janus colloids in an evaporating sessile droplet with 10\% hydrogen peroxide, under an atmosphere of 75\% relative humidity, after 2000 seconds. Particle positions are obtained in three dimensions using their defocused image, and the droplet geometry is obtained from side view imaging.}
	\label{fig:1}
\end{figure}

A significant body of research has been devoted to active Janus particles trapped at interfaces: Nobili, Stocco \emph{et al.} \cite{WangStocco2015interface,WangStocco2016interface,WangStocco2017interface,stocco2019rotational} have carefully analyzed how the pinning of the interface at the particle's surface constrains their rotational diffusion, enhancing their ballistic regime and their average propelling velocity. Active Janus colloids at liquid-liquid interfaces have been shown to be strongly propelled by a self-induced Marangoni effect, becoming ``Marangoni surfers'' \cite{dietrich2020microscale} when illuminated with a locally strong laser beam. Even for non-trapped active phoretic particles, the proximity of a free interface does influence the particle dynamics \cite{malgaretti2021phoretic}. While in the aforementioned studies active Janus particles have purposely been placed at interfaces, few have studied the way particles interact with the interface from the bulk, where they are typically dispersed. Apart from a few exceptions \cite{simmchen2017active,kellay2022droplet}, the number of experiments with active matter in droplets is scarce, due to the natural three-dimensional character of the system, the presence of dynamic interfaces and other complications ascribed to evaporation.

In this paper, we explore the dynamics of Janus catalytic colloidal particles in evaporating droplets of a dilute hydrogen peroxide solution. Our motivation is to identify those phenomena that could be found in any other active phoretic colloidal entity, biological or synthetic, within a sessile droplet. Evaporating droplets can typically manifest in a wide range of evaporation-driven flows \cite{Deegan1997Nature,Lohse2022ARFM,Gelderblom2022Review}, with an intensity being inversely proportional to the relative humidity of the atmosphere surrounding the droplet. Apart from the motion generated in liquid phase by the evaporation, active Janus catalytic particles are self-propelled. In this paper, we make use of 3-\um-diameter fluorescent and sulfate-functionalized polystyrene particles (PS-Fluored-3 from Microparticles GmBH) coated with platinum on one hemisphere. Such a configuration induces an asymmetric reaction of hydrogen peroxide in the solution into water and oxygen only on the catalytic active side. This results in chemical \cite{Howse:2007ed} and/or ionic \cite{BrownPoon2014ioniceffects} gradients from hemisphere to hemisphere that drive a flow towards the catalytic side of the particle by chemical and/or ionic diffusiophoresis \cite{BrownPoon2014ioniceffects,Ebbens2014ionic}, propelling the particle towards the non-coated side (see Fig. \ref{fig:1}).

We will show experiments of a diluted population of active Janus particles in droplets of water and hydrogen peroxide (initial concentration 10\%) in controlled atmospheric conditions and varying relative humidities: from saturated conditions (100\% relative humidity, section \ref{sec:noevapor}) to unsaturated  (50\% relative humidity), in section \ref{sec:evapor}. Additional control experiments are performed with passive colloids, which are typically used as flow tracers to infer the convective flow \cite{Rossi:2019,marin2011order,Gelderblom2022Review}. The experimental results are contrasted with active Brownian particle simulations and finite element simulations for the fluid flow.

The paper is organized as follows: After describing our methodology in section \ref{sec:methods}, we show experiments performed in the absence of evaporation that evidence significant agglomeration at the interface in section \ref{sec:noevapor}. We finalize that section with active Brownian dynamics simulations that help us interpret the experimental results. In section \ref{sec:evapor}, we discuss the consequences of the presence of an evaporation-driven flow in the dynamics of the active particles using passive particle tracers and finite difference simulations. The paper ends with an overview of the results and comparisons with similar systems in section \ref{sec:conclusions}.

\section{Experimental Methods}\label{sec:methods}

{Our Janus catalytic particles are fabricated from a solution of 2.75 \um~red fluorescent polystyrene particles (Microparticles GmbH) at 1 wt\%. The solution is spin coated onto a glass microscope slide, previously cleaned with acetone and plasma-cleaned, and coated with a nanometric film of platinum (approximately 17 nm thick) using an e-beam evaporation system (in-house system built in Univ. Twente's NanoLab, based on a Balzers BAK system). After the evaporation process, the particles are retrieved by immersing the slide in water and sonicating it for 25 min. The resulting diluted suspension is then centrifuged, supernatant removed, and a solution of approximately 1\% wt is obtained.}

\begin{figure}[t]
	\centering
	\includegraphics[width=0.95\textwidth]{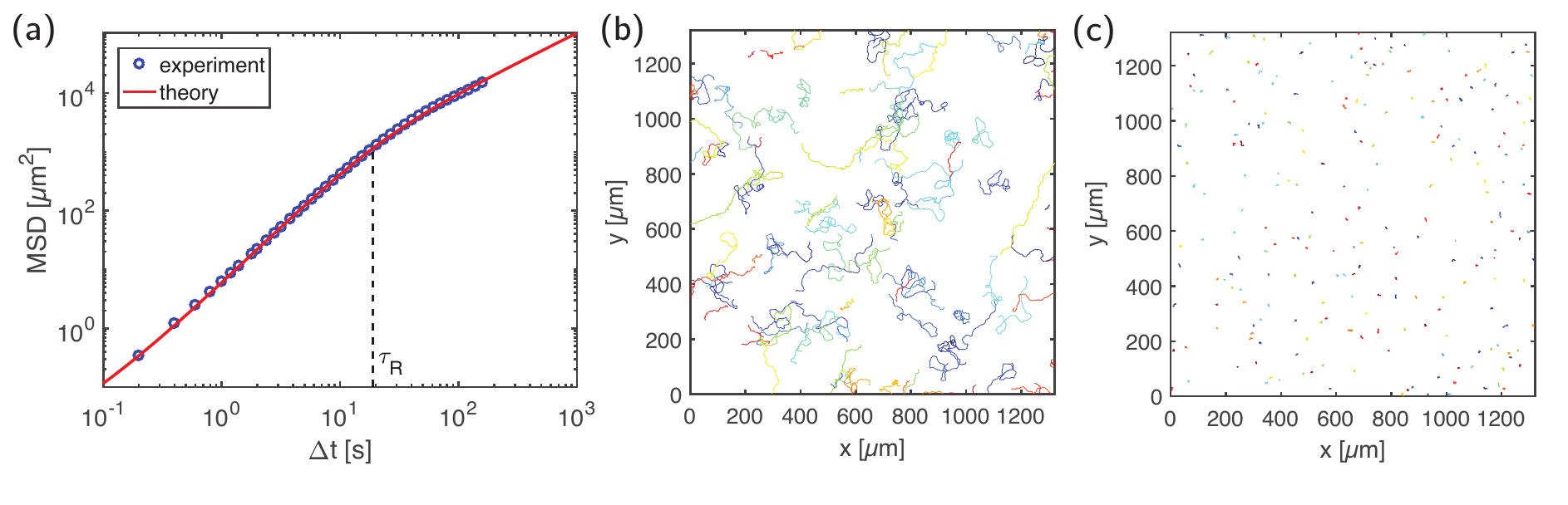}
	\caption{(a) Average mean squared displacement (MSD) of Janus catalytic particle solution in the bulk of a 15\% \perox~ solution. Fitting the parameters of the model in Eq. \ref{eq:MSD} yields $v_p=$ 2.9 $\mathrm{\mu m}$/s and $\tau_R=$ 19 s.
	(b) A selection of the active Janus particles trajectories used for assessing their self-propulsion velocity and rotational diffusion. {These trajectories are obtained close to the bottom of a large container. Only the XY positions are taken into account to compute the MSD shown in (a).}
	{(c) When the Janus particles are suspended in water, they tend to sediment to the bottom of the container, and there the only motion observed is Brownian, as shown in the plot. The trajectories shown here have been collected for about 5 min, the same time as in (b).}}

	\label{fig:MSD}
\end{figure}

To test the Janus particles' activity, the previous solution is mixed with hydrogen peroxide at 15\% wt and sonicated for a few minutes. 1 $\mathrm{\mu}$L of that solution is deposited in a closed cuvette with a thin transparent bottom surface, and observed using an inverted microscope. The chamber allows visual access for (1) an inverted microscope to the inside of the droplet, and (2) a side view of the droplet profile, from which we will obtain the droplet volume as a function of time. Focusing in a layer far enough from the upper surface, we measure the two-dimensional trajectories of hundreds of particles (see Fig. \ref{fig:MSD}b), which will be analyzed to obtain their mean squared displacements (MSDs). The average MSD is shown in Fig. \ref{fig:MSD}a. It is expected to contain a diffusive term and a propulsive term, following \cite{TenHagen2011}

\begin{equation}\label{eq:MSD}
\left<(\Delta r)^2\right> = 4D_T\Delta t + \frac{V_p}{3D_R^2}[2D_R\Delta t+e^{-2D_R\Delta t}-1],
\end{equation}    
    
\noindent where $D_T$ is the translational diffusion of the colloidal particles $D_T=k_BT/6\pi\mu a$, $k_BT$ is the thermal energy, $\mu$ the liquid dynamic viscosity and $a$ the particle radius. $D_R$ and $V_p$ are respectively the rotational diffusion coefficient and the particle's propulsion velocity, which are obtained by fitting the experimental data to this model (see Fig. \ref{fig:MSD}a).

To obtain the distribution of the particles within the droplet volume, we track their positions a few seconds after the droplet has been deposited. Their three-dimensional position is obtained through the defocused image of the microparticles using the open software \emph{defocustracker}, a modular and open-source toolbox for defocusing-based 3D particle tracking developed by Rossi \& Barnkob \cite{barnkob2020general,barnkob2021Defocus}. 
Particle images are acquired using an sCMOS PCO Edge 5.5 camera with 2560$\times$2160 pixel resolution, coupled with a Nikon Eclipse Ti2 inverted microscope using a 10$\times$ objective, which yields a spatial resolution of 0.62 \um/pixel. This spatial resolution has been carefully chosen reaching a compromise between resolution and target volume. A similar compromise needs to be made with respect to the time resolution: A recording rate of 3.88 frames per second is chosen, which allows for individual experiments of 600 seconds (due to the limited storage memory of the camera), and enough temporal resolution to capture the particle's motion with accuracy.
Particle positions in the optical axis can be obtained with \emph{defocustracker}\cite{barnkob2021Defocus} with errors in the order of one particle radius. The distance between particle and the moving interface is obtained indirectly by combining the particle's bottom view and the droplet's side view, and it is obtained with an absolute error that accumulates up to two particle diameters in the worst cases. Unfortunately, this means that our experimental approach does not allow to obtain the relative distance of the particle with the interface with sufficient precision to conclude whether the particle is adsorbed at the interface, or only close to it. Nevertheless, as we will discuss below, we are able to reach conclusions on whether particles are adsorbed or not, but based instead on the interface-particle relative motion.

\section{Accumulation in the absence of evaporation}\label{sec:noevapor}

\begin{figure}[t]
	\centering

	\includegraphics[width=0.95\columnwidth]{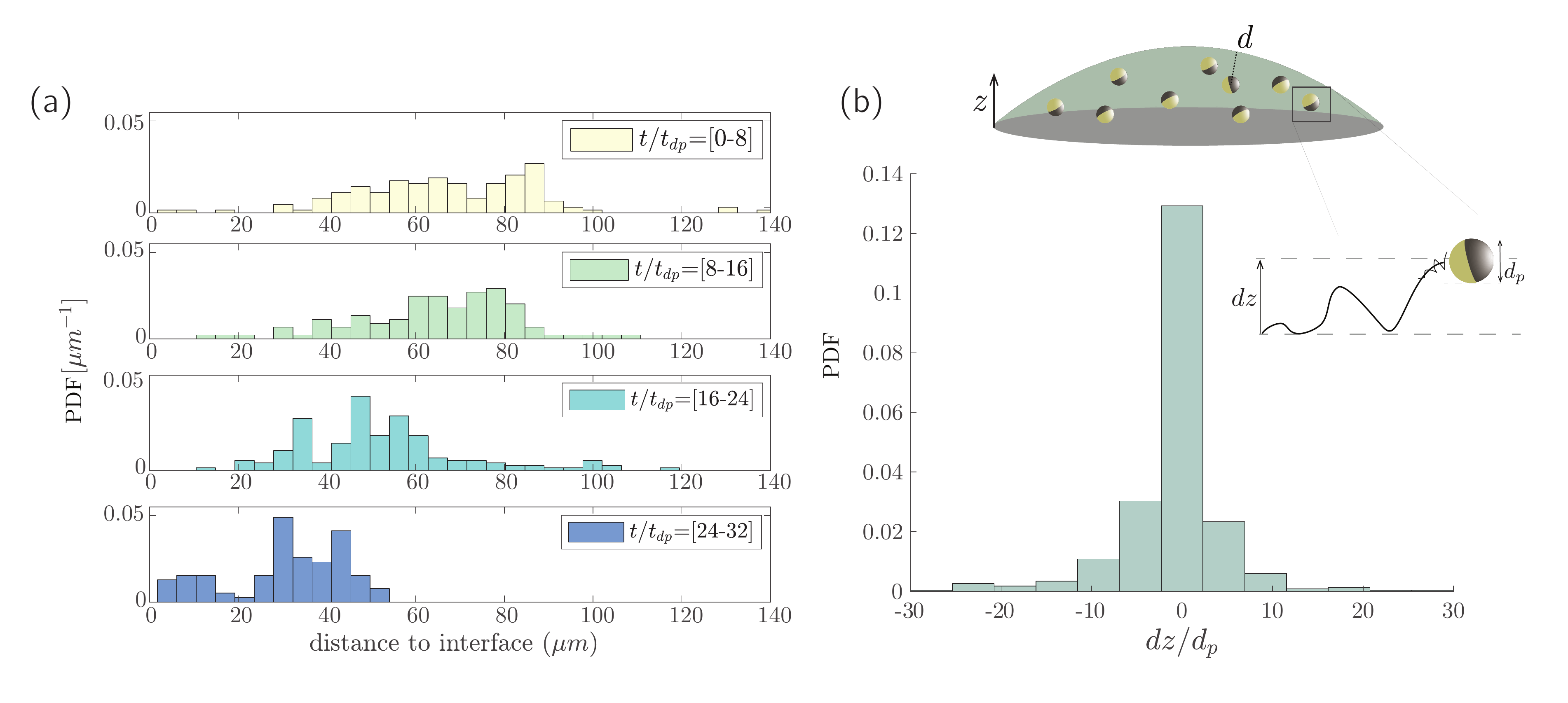}
	\caption{(a) Distribution of particle average distances to the liquid/air interface in a 40 minutes experiment in the absence of evaporation. The shift of the distribution to the left side is a clear sign of the accumulation of active Janus particles in the vicinity of the interface. (b) The distribution of self-propelled particle's vertical displacements (in the gravity axis) is symmetrical and centered at $dz=0$. Janus active particles in this size range are not observed to suffer from gravitaxis and consequently it cannot be invoked to explain the interfacial particle accumulation in our results.}
	\label{fig:noevaporPDF}
\end{figure}

In the absence of evaporation, the droplet remains practically unperturbed for long periods of time, allowing to track particles trajectories under identical conditions in several consecutive recordings. The particle positions and the average distance of each particle to the droplet liquid/air interface were computed for an experiment running over 40 minutes (4 consecutive recordings). In Fig. \ref{fig:noevaporPDF}a, we show the evolution of the distribution of particle distances to the interface. In the first 10 minutes (first panel in Fig. \ref{fig:noevaporPDF}a), the particles are mostly concentrated in the bulk of the droplet as expected. Surprisingly, as time evolves, they seem to slowly ``migrate'' from the bulk of the droplet to the vicinity of the liquid/air interface. Control experiments with the Janus particles in the absence of \perox~ have also been performed with the expected results: after sedimenting to the bottom of the droplet, the only motion detected on the particles is Brownian. Additional control experiments with neutrally buoyant passive particles in the presence of \perox~ in the saturated environment showed no signs of evaporation-driven flow.
It is important to mention that the bottom surface is not visible in this particular example shown in Fig. \ref{fig:noevaporPDF}a, but a smaller population of particles can also be found at late stages in the vicinity of the bottom solid surface (see Fig. S2 in the Supplementary Material at [URL will be inserted by
publisher] for a different example).

Although our tracking technique does not allow to obtain the relative particle position to the interface with enough precision (as discussed in section \ref{sec:methods}), we can clearly infer that most particles have not breached the liquid-air interface: interfacial trapping clearly does not occur in our case since not a single particle has been observed to remain swimming along the droplet's surface. Instead, particles typically swim in and out of the vicinity of the interface. This is probably due to the hydrophilic coating on the polymer side, which prevents the particles to breach spontaneously the interface \cite{kaz2012physical}. Interestingly, this behavior can be reproduced in active Brownian particle simulations with simple but generic boundary conditions as we will discuss further below.

Campbell et al. \cite{Campbell2017} recently observed that Janus catalytic particles could experience gravitaxis due to the torque exerted by gravity on the heavier catalytic side. This results in a net upwards drift of the particles at longer times which could be responsible for the ``migration'' just shown. To test whether gravitaxis plays a role in our system, we computed the total distance traveled in the direction of gravity $dz$ per particle, and plotted its distribution of values, where $dz<0$ indicates the direction of gravitational acceleration. Figure \ref{fig:noevaporPDF}b clearly shows that the particle displacement distribution is well-centered around $dz=0$, from which we interpret that our particles are either too small, or their platinum coating too thin, to manifest gravitaxis \cite{Campbell2017}.

\begin{figure}[h]
	\centering
	\includegraphics[width=0.95\textwidth]{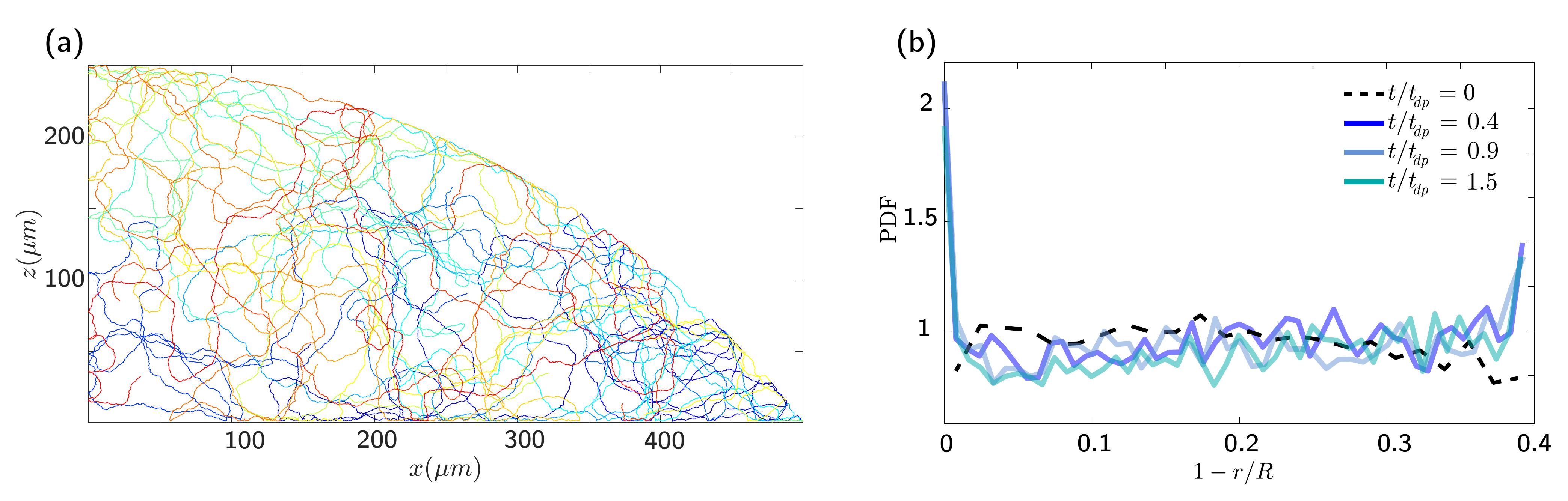}
	\caption{(a) A random selection of numerically computed trajectories of self-propelled particles are shown within a confined boundary resembling a droplet. The droplet interface is represented by quarter of a circle centered at $r=0$ (not shown in the Figure) and with a radius $r=R$. (b) The graph shows the probability distribution as a function of the distance to the interface ($1-r/R$) of the particles at four instants of the simulation: initially ($t=0$) particles are randomly distributed. Shortly after the simulation starts certain interfacial accumulation can be observed at $r=R$, which keeps increasing until the end of the simulation, {at $t_{end} = 1.5 t_{dp}$, where $t_{dp}$ is the time taken by a phoretic particle to travel one droplet radius}. Certain accumulation at the solid substrate bottom surface can be also observed.}
	\label{fig:sims}
\end{figure}

In order to further understand the mechanism behind the accumulation of particles in the vicinity of the liquid/air interface in the absence of evaporation, we devise a simple numerical two-dimensional model consisting of active particles inside a confined system of comparable cross-section as one of our droplets. The active particle motion is solved in 2D by numerically integrating a set of Langevin equations \cite{Howse:2007ed,TenHagen2011,Volpe2014}, in which the particle self-propelling velocity is given as an input (obtained from our own particles $V_p\approx 3$ \um/s. 
In this numerical analogue, 10000 non-interacting particles are introduced in an area as shown in Fig. \ref{fig:sims}a, limited by three boundaries in which we impose different boundary conditions. First, at the droplet symmetry axis ($x=0$), we apply a periodic boundary condition. Secondly, at the liquid-gas interface ($r=R$), we apply a non-reflective condition, i.e. particles attempting to cross the interface are allowed to remain at the interface, with their velocity component normal to the boundary neutralized. No additional forces are added to emulate the interaction with the free interface. Finally, at the bottom surface ($z=0$), we apply two boundary conditions: a non-penetration condition and a normal lubrication force. The latter consists of a force activated when a particle is at one particle diameter from the surface, directly proportional to the velocity component perpendicular to the surface and inversely proportional to the distance to the surface (see Section V in the Supplementary Material at [URL will be inserted by
publisher] for more details). In Fig. \ref{fig:sims}b we plot the probability distribution of the particles in this simple numerical model, plotted as a function of $1-r/R$, where $R$ is the circular segment's radius (not the droplet radius), $r$ is the distance from the circle's center (not the droplet center), and $1-r/R=0$ corresponds to the droplet's interface $r=R$. The probability is normalized by the area of the contained in every radial bin.

Time is normalized based on the time taken by a particle to travel a distance equivalent to a droplet radius $t_{dp}\approx 345$ s. At $t=0$ the particles are randomly distributed in the volume (dashed black line in Fig. \ref{fig:sims}b). After only 0.4 $t_{dp}$, a significant accumulation can be observed at $r/R =1$. This interfacial accumulation is kept steady up to $t = 1.5 t_{dp}$ (same order of magnitude as a typical experiment).Analyzing the trajectories, one can observe that a significant amount of particles are observed to swim along the interface for long periods of time, similar to what they do in the experiments. In addition, a non-negligible amount of particles also are observed to remain in the vicinity of the solid surface, but in a lower amount compared with the interface. 
Nonetheless, a significant --and interesting-- quantitative difference is found in the time taken for the particles to populate the interface. While in experiments particles only start to populate the interface after $t/t_{dp}\approx$10, in the simulations they do after $t/t_{dp}\approx$0.4. This significant mismatch is most likely due to the dimensional reduction in the simulations, as particle arrive sooner at the interface in two-dimensions.

 We therefore can conclude that, while other mechanisms might enter into play in the experiments (e.g. hydrodynamic interactions with the free interface), the driving mechanism leading to the accumulation of particles at the liquid-air interface is their dominant self-propelled nature over thermal fluctuations \cite{malgaretti2021phoretic} (clearly quantified by a P\'eclet number $Pe = V_p r_p/D_T \approx 24$). 
 {Indeed, other phenomena might be present like long range hydrodynamic interactions \cite{BergLauga2008,Lopez_Lauga2014} or the presence of ``sliding states'' due to a combination of hydrodynamic and chemical interactions with boundaries \cite{uspal2015self}. However, such mechanisms have only been shown to be present at liquid-solid surfaces, it would need to be shown that they apply for liquid-liquid cases.}

{ A similar behavior has been observed both for \emph{E. coli}\cite{vladescu2014ecolidroplet} and for active Janus particles in highly confined liquid droplets\cite{kellay2022droplet}. }In those cases, the typical mean free path of the swimming micro-entities was larger than the confinement, analogously to a tightly confined rarefied gas. This is not our case, since the confinement space is much larger, but nevertheless an interfacial accumulation is quickly observed, which suggests that the mechanism proposed here might be present in other systems. For example, \emph{E. coli} and \emph{P. aeruginosa} have often been observed to swim along liquid interfaces \cite{VaccariStebe2017bacteria,vaccariStebe2018cargo}, and eventually even get trapped at the interfaces.

\section{Active Janus particles in sessile droplets in the presence of evaporation}\label{sec:evapor}

\begin{figure}[h]
	\centering
	\includegraphics[width=0.95\textwidth]{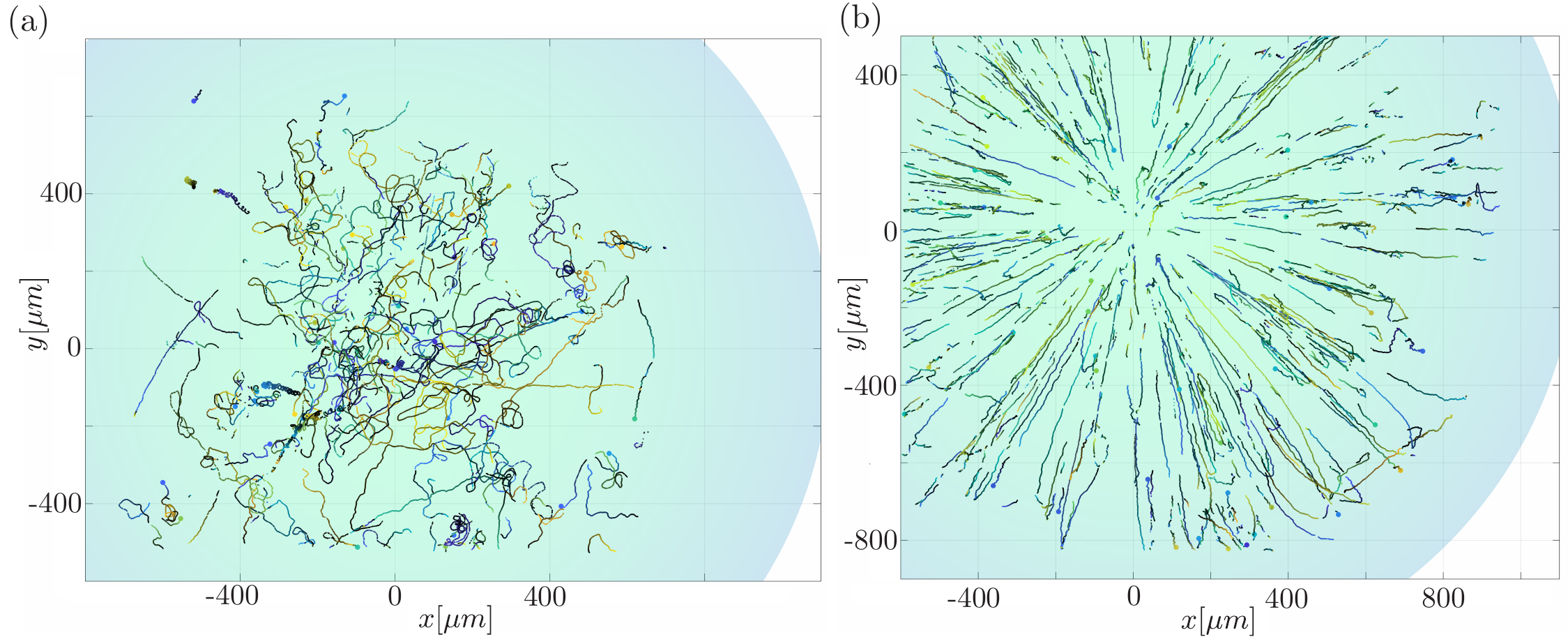}	
	\caption{Active and passive colloids in an evaporating droplet under identical conditions: (a) Trajectories of active Janus colloids, 10\% \perox~ and RH=75\% (b) Trajectories of passive colloids at the same conditions, 10\% \perox~ and RH=75\%. }
	\label{fig:trackscomp}
\end{figure}

Allowing the sessile droplets to evaporate introduces a number of new features. First of all, from the technical point of view, the system evolves now in time, which sets a limit on the amount of data that can be captured in every experiment.
Secondly, since pure hydrogen peroxide has larger surface tension than water (80.2 mN/m at 20$^\circ$C \cite{phibbs1951hydrogen}, compared to 72.8 mN/m for water under the same conditions) and lower vapor pressure (approx. 300 Pa at 25$^\circ$C for \perox, compared to 3.1 kPa for water at the same temperature), we expect water to evaporate faster in a sessile droplet, leaving the contact lines enriched with \perox , and generating a surface tension gradient directed towards the contact line. To confirm this, we performed both experiments using passive colloids and as well as finite element simulations that confirm our hypothesis (see Supplementary Material at [URL will be inserted by publisher] for the finite element simulations in the fluid flow). The internal flow within the droplet is practically dominated by an interfacial Marangoni flow, directed from the droplet's apex to its contact line. Such a flow is directly proportional to the relative humidity imposed in the chamber \cite{Gelderblom2022Review}. In a similar way as the experiments in Section \ref{sec:noevapor}, we performed experiments at RH=75\% and RH=50\%, using both active Janus particles at 10\% initial concentration of \perox~ as in previous sections. 

It is important to note that a new time scale enters into play when the droplet evaporates at a non-negligible rate, which is the droplet evaporation time. The order of magnitude for the life time of a diffusion-limited evaporating droplet is given by $t_D=\rho R_D^2/2D_v\Delta{C}$, where $\rho$ is the liquid's density, $R_D$ the droplet's contact radius, $D_v$ is the vapor's diffusion coefficient, and $\Delta C=C_S(1-RH)$ is the vapor concentration difference from the droplet's interface (at $C=C_S$) to the atmosphere (at $C=RH\times C_S$). Although our droplets are $\mathrm{H_2O}$/\perox~ mixtures, we will assume that the main contribution to the evaporation during our experiments is due to the most volatile phase, i.e. water, and therefore we will make use of the droplet's life time estimation as if it were a pure water droplet. This is an important point, since the duration of our experiments is limited by the camera's limited memory. Therefore, in experiments with RH=75\%, $t_D\approx2174$ s, and the experiments shown here capture a time duration ratio $t_{exp}/t_D=0.26$, and $t_{exp}/t_D=0.52$, in the case of RH=50\%.

Starting with the results for RH=75\%, we performed experiments with both, a diluted solution of passive colloidal particles (polystyrene red-fluorescent 1-\um-diameter particles) and with the solution of Janus active particles. The trajectories of each type of particle are shown in Fig. \ref{fig:trackscomp}. The x-y-projected trajectories of the active Janus particles resemble those obtained under vapor saturated conditions in Fig. \ref{fig:1} and also \ref{fig:MSD}. In contrast, the trajectories of the passive colloids clearly evidences the presence of an evaporation-driven flow directed radially outwards from the droplet's apex, qualitatively identical as our finite element simulations, which can be found in the Supplementary Material at [URL will be inserted by publisher].

\begin{figure}[t]
	\centering

	\includegraphics[width=0.85\textwidth]{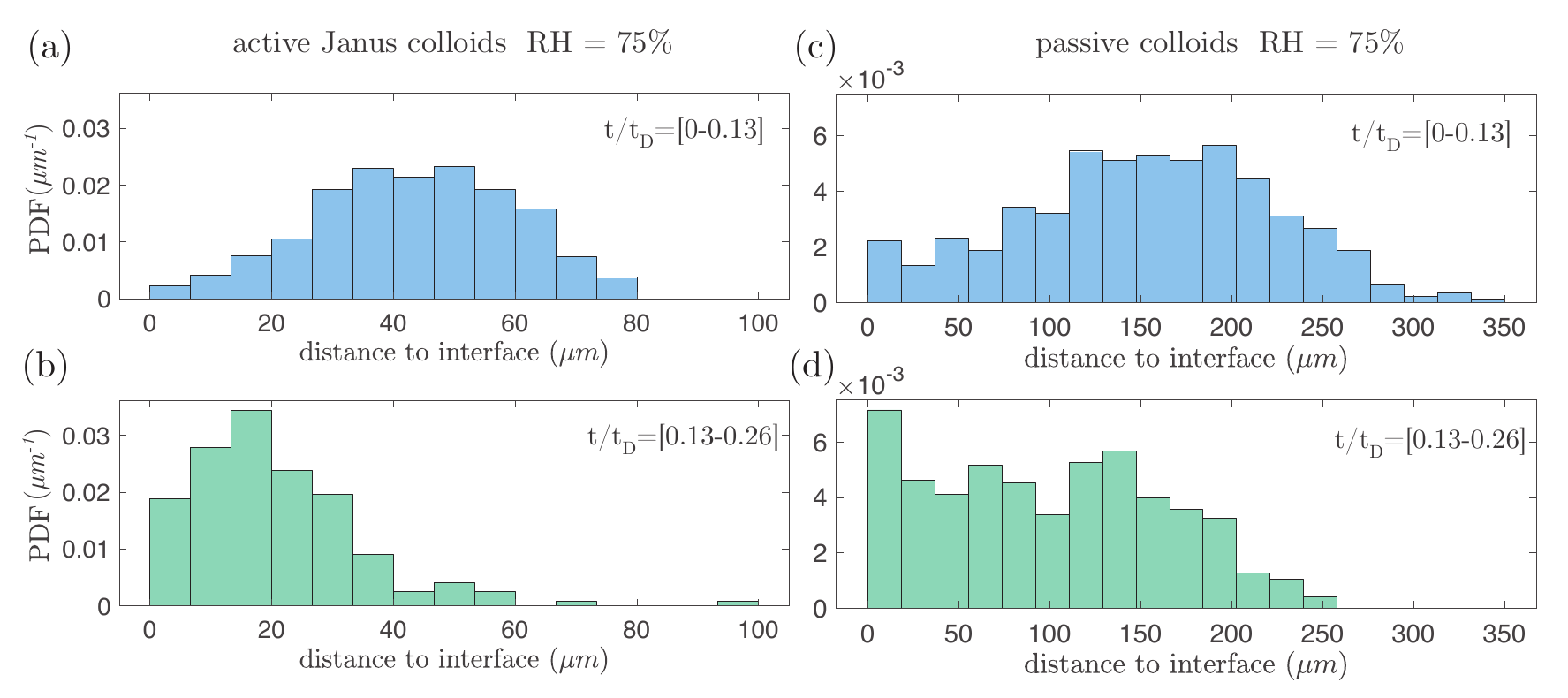}
	
	\caption{Distribution of particle distances to the liquid/air interface in evaporating droplets at RH=75\%. (a) Active Janus colloids in the first period of the evaporation process (from $t=0$ s up to $t/t_D =$ 0.13, corresponding to $t/t_{dp}=$0.83). Newly deposited droplets typically fairly homogeneous bulk concentration with particle-depleted interfaces, similarly with passive colloids in (b). (c) Active Janus colloids accumulate notably at the droplet's interface in only $t/t_D=$0.26. Interestingly also passive colloids do tend to accumulate at the interface (see (d)) but to a lesser degree. The receding interface in evaporating droplets explains the accumulation in the passive colloids and the stronger rate in the active ones.}
	\label{fig:pdfcomp}
\end{figure}

On this occasion, the distribution of particle positions of both passive and active colloidal particles are analyzed. Figures \ref{fig:pdfcomp}a and b show the active particles distribution as a function of the distance to the interface in the first half of the experiment (Fig. \ref{fig:pdfcomp}a, up to $t/t_D=0.13$), and on the second half (Fig. \ref{fig:pdfcomp}b, up to $t/t_D=0.26$). The accumulation of particles in the interface vicinity is again evident, and actually even stronger than in the absence of evaporation. After $t/t_{dp}=$1.6, most particles are found in about 40 \um~ from the interface in the presence of evaporation, while it takes more than $t/t_{dp}=$5 to reach a similar distribution in experiments in the absence of evaporation.
Figures \ref{fig:pdfcomp}b and \ref{fig:pdfcomp}d show the distribution of passive colloids, which deserves to be analyzed carefully. First of all, we should note that the range of distances to the interface that we can track is larger than with Janus colloids ($\Delta z$ up to 300 \um, compared to 150 \um~for this particular example of active Janus particles). This is due to the reduced fluorescence emission of the Janus particles due to the presence of the platinum coating. Then, we can clearly see that at the end of the experiment ($t_{exp}/t_D=0.26$), the bulk of the droplet is slightly depleted and the vicinity of the interface enriched. This phenomenon is observed in evaporating suspension droplets for most colloidal particles and can be explained by the apparent receding motion of the free interface towards the particles \cite{trantum2013cross,Jafari2016,Rossi:2019,Gelderblom2022Review}, which \emph{sweeps} particles as the droplet evaporates and its interface seems to recede. Note that the interface does not strictly ``move''; liquid is being evaporated from it, and therefore there is no bulk motion caused by the evacuation of liquid from the interface, in that sense, the receding interfacial ``motion'' is apparent. This effect explains the interfacial enrichment of particles observed for passive colloids in our experiment, which keeps increasing as the droplet evaporates \cite{Gelderblom2022Review}.

\begin{figure}[h]
	\centering
	\includegraphics[width=0.9\textwidth]{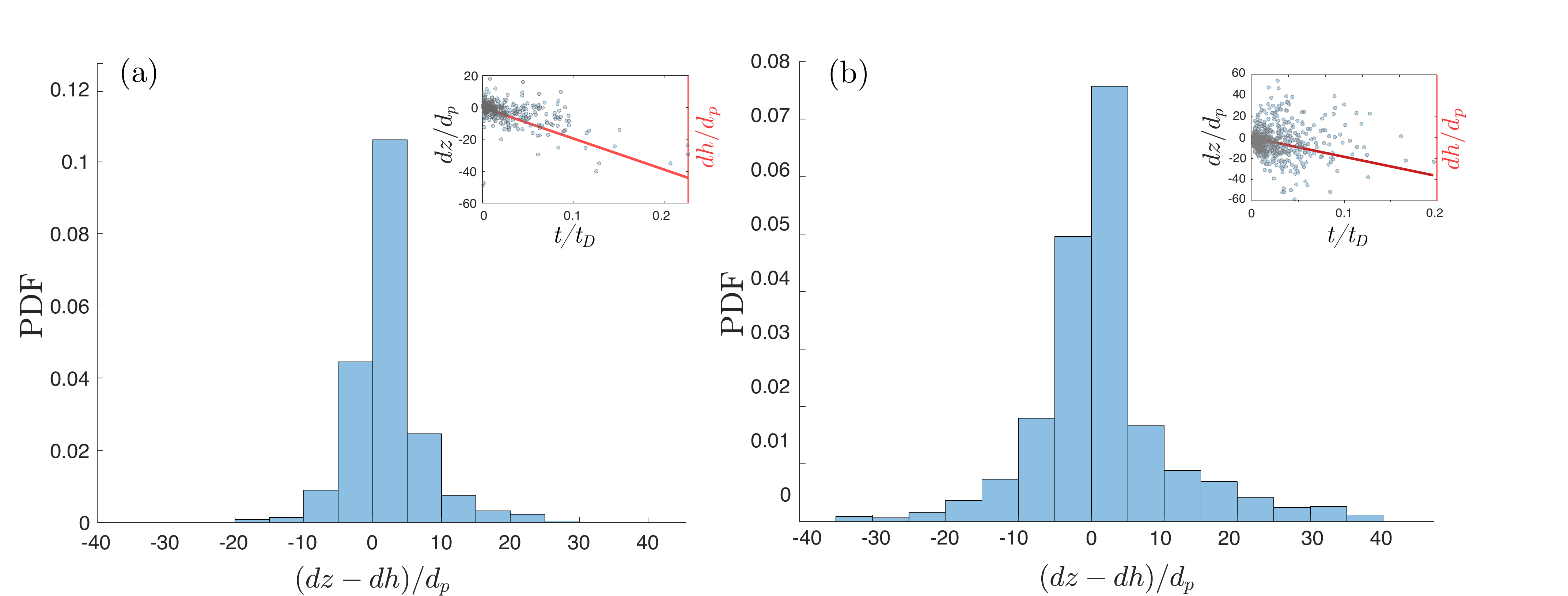}
	\caption{{Probability distributions of $dz-dh$, where $dz$ is the distance traveled in the gravity axis by (a) an active Janus particle and (b) a passive colloidal tracer within its lifetime. $dh$ is the distance covered by the droplet's apex while evaporating in an environment at RH = 75\%. In the inset one can see that $dz$ and $dh$ are correlated by plotting both independently against the particle lifetime $t/t_{st}$, with $t_{st}$ being the time taken for a particle to travel its own diameter. Janus active particles, due to their vicinity to the liquid-gas interface, manifest a stronger correlation than passive colloids, which tend to remain recirculating within the droplet's volume dragged by the evaporation-driven flow.}}
	\label{fig:dzdh}
\end{figure}

Interestingly, such an ``interfacial sweeping'' mechanism must also be present with Janus particles in evaporating droplets, which adds up to the previously discussed mechanism in the absence of evaporation, based on the particle's self-propulsion. To test this hypothesis, we compute again the vertical distance traveled by every particle in its lifetime $dz$, as we did in section \ref{sec:noevapor}. To quantify the apparent receding motion of the interface, we simply compute experimentally the droplet's apex receding velocity $\Delta h/\Delta t$ and, since this is constant for diffusion-limited evaporating droplets with pinned contact lines \cite{cazabat2010evaporation,Gelderblom2022Review}, from this we compute the distance $dh$ that the interface recedes during the lifetime of every active particle. As shown in the inset of Fig. \ref{fig:dzdh}, $dz$ and $dh$ are fairly well correlated with each other, and to further show this, we compute the histogram of $(dz-dh)/d_p$, which is well-centered around $(dz-dh)/d_p=0$ and decaying to negligible values at 10 particle diameters. Note that this histogram is practically identical to that in Fig. \ref{fig:noevaporPDF}b, as it is expected.
{On the contrary (and somehow paradoxically), passive tracer colloids tend to explore a much larger volume of the droplet than active colloids in an evaporating droplet. This is because, dragged by the solutal Marangoni flow, they circulate in streamlines that cover a significant volume of the sessile droplet. For this reason, each passive tracer explores a larger range of $dz$ values, which is shown in a wider histogram in Fig. \ref{fig:dzdh}b and in its inset. Nevertheless, the fact that the histogram of $dz-dh$ is fairly well-centered around 0 in Fig. \ref{fig:dzdh}b is a sign that in the long-term, passive particles are also subjected to the interfacial sweeping.}
We can conclude here that the ``interfacial sweeping'' indeed contributes to the interfacial agglomeration of particles, but it is not the only mechanism leading Janus active particles to the interface vicinity.

By reducing the relative humidity down to RH=50\%, we also increase the speed of the evaporation-induced flow. For example, the typical velocity of the convective flow in RH = 75\% (measured through the median passive particle velocity) is about 1.4 \um/s, half of the typical propelling velocity of an active particle ($V_p\approx 3$ \um/s). In the case of RH=50\%, the convective flow velocity increases up to 4 \um/s and the role of the particle's activity is then shadowed and clearly dominated by the convective flow. Consequently, the interfacial accumulation phenomenon is expected to be caused mainly by the ``interfacial sweeping'' mechanism as the relative humidity is further reduced (results shown in the Supplementary Material at [URL will be inserted by publisher]). 

In this context, a dimensionless number could be defined for Janus catalytic particles as the ratio between their self-propelled velocity $V_p$ (measured through the MSD's ballistic regime) and the convective flow velocity $V_f$, $\Upsilon = V_p/V_f$. Such a propulsion-convection number would give us essentially infinite values for the non-evaporative case (since no significant convective flow could be measured). One must be aware nonetheless that evaporating-driven flows are naturally heterogeneous and unsteady in time, and therefore such propulsion-convection number $\Upsilon$ would vary when assessed locally and instantaneously. Since we are using global averages to characterize the particle velocities, we could define a $\left<\Upsilon\right> =$ 2.07 for RH=75\%, where the self-propulsion is still dominant, and $\left<\Upsilon\right>= $ 0.73 for RH=50\%, a value at which the self-propulsion is still visible, but the trajectories are strongly influenced by the convective flow.

\section{Conclusion}\label{sec:conclusions}

In this work we have analyzed the three-dimensional trajectories of active Janus catalytic particles in evaporating sessile droplets of water and hydrogen peroxide for different relative humidities: from RH$\approx100$\%, where no evaporation-driven flow can be detected, down to RH=50\%, where strong Marangoni flows are observed.

In the absence of evaporation, Janus catalytic colloids tend to accumulate in the droplet's interface vicinity, driven by their own self-propelled motion, where they remain for most of the period observed (up to 2400 Stokes times or 6.5$\times t_{dp}$). A smaller population of particles is also observed to accumulate at the bottom's solid surface. A simple numerical model of active Brownian particles in a 2D system is employed to verify that such a phenomenon is inherent to any system with phoretic particles and free interfaces, even at moderate confinement. The numerical system has a free interface, where only a no-penetration condition is imposed, and a boundary at the bottom with a no-penetration condition and a normal lubrication force, proportional and opposite to the particle's normal velocity. The simulations show a high probability of finding particles at the boundaries, with a significantly higher population at the free interface, just as observed in the experiments. 

We propose that this interfacial accumulation mechanism could be generic for pusher squimmers in moderate confinement. Self-propelled Janus particles in this size range do not experience significant gravitaxis, interacting with interfaces in a similar way as biological swimmers do, and therefore remain being good candidates as model active system \cite{Poon2016Ecoli}. 
An important point is the interaction of the particles with the interface. Our particles are observed to remain swimming along the interface, but they do not get trapped at it (as \emph{E. coli} or \emph{P. aeruginosa} often do). A reason could be the sulfate coating that conveys stability to the polystyrene particles solution, and which conversely also conveys certain repulsion from the solid bottom surface (borosilicate glass). 
{Interestingly, in the recent work of Bitterman \emph{et al.}\cite{Deblais2021algae} the authors observed certain accumulation of the microalgae \emph{C. reinhardtii} at the border of evaporating sessile droplets, probably experiencing a similar phenomenon as here observed for synthetic particles.}

When active Janus particles are trapped at interfaces \cite{WangStocco2015interface,WangStocco2016interface}, their trajectories are strongly influenced by the constrained rotation caused by the pinning of the liquid's interface to their solid surface, and consequently, their rotational diffusion tends to be significantly reduced \cite{stocco2019rotational}. Although trapping has not been observed in our experiments, the proximity to the free interface should have an effect on their trajectories \cite{malgaretti2021phoretic} that still needs to be further investigated. 

As soon as the water/hydrogen peroxide droplets are allowed to evaporate, the liquid's interface seems to recede, dragging particles along in the interface's vicinity in a ``sweeping'' effect. This phenomenon is well-known to occur with passive colloids \cite{trantum2013cross,Jafari2016,Rossi:2019}, and it does also contribute to a stronger interfacial accumulation of active Janus particles at the interface's vicinity. The droplet's evaporation also triggers an important phenomenon: an interfacial Marangoni flow appears, directed from the droplet's apex towards the contact line. Such a flow does not have a significant effect on the trajectories at moderate evaporation rates (RH=75\%), which we propose to quantify using a propulsion-convection number $\Upsilon=V_p/V_f$, which takes an average value of $\left<\Upsilon\right>=$2.03. By increasing the evaporation rate (RH=50\%), the convective flow increases, and the propulsion-convection number reduces to $\left<\Upsilon\right>=$0.73, the active Janus particles trajectories are clearly influenced by the presence of the flow. Nonetheless, due to the strong non-homogeneous and unsteady character of evaporation-driven flows, their influence on self-propelled particles within sessile evaporating droplets needs to be further investigated, especially in the vicinity of contact lines \cite{Deegan1997Nature,marin2011order}. 

Most of the phenomena observed here for Janus catalytic active colloids should also be present in other self-propelled micro-entities inside sessile droplets: interfacial accumulation driven by self-propulsion (even in the absence of strong confinement as in references \cite{vladescu2014ecolidroplet,kellay2022droplet}), the ``interfacial sweeping '' effect when evaporation is present and the influence of the convective flow in the trajectories as the evaporation rate increases. At very high evaporation rates, the role of self-propulsion is expected to be rather minor ($\left<\Upsilon\right>\ll1$) and the micro-entities within the droplet might behave as passive colloids if their swimming power is not sufficient. However, interactions with the free interface, heterogeneities within the droplet's volume and unsteadiness are crucial points that would still need to be carefully addressed in the future.

\begin{acknowledgments}
AM \& DL both acknowledge funding from the European Research Council, with grant numbers 678573 and 740479 respectively, and the Max-Planck Center Twente, cofinanced by NWO. BtH gratefully acknowledges financial support through a Postdoctoral Research Fellowship from the Deutsche Forschungsgemeinschaft – HA 8020/1-1. MJ, BtH and AM, acknowledge fruitful conversations with Hartmut L\"owen and Jan Eijkel.
\end{acknowledgments}




%

\end{document}